\documentclass[pra,twocolumn,superscriptaddress,floatfix,amsmath,nofootinbib,amssymb]{revtex4-1}

\usepackage{graphicx}
\usepackage{bm}
\usepackage{verbatim}
\usepackage{mathrsfs}
\usepackage{color}
\usepackage{paralist}
\usepackage{hyperref}
\usepackage{subfigure}

\newcommand{\ii}{\mathrm{i}}

\newcommand{\ket}[1]{| {#1} \rangle}

\begin{document}

\title{Taming an Optical Schroedinger's cat - A quantum non-demolition approach}
\author{Marvellous Onuma-Kalu}
\email{monumaka@uwaterloo.ca}
\affiliation{Department of Physics \& Astronomy, University of Waterloo,  Ontario Canada N2L 3G1}
\author{Kae Nemoto}
\affiliation{National Institute of Informatics, 2-1-2 Hitotsubashi, Chiyoda-Ku, Tokyo 101-8430, Japan}
\author{W. J.  Munro}
\affiliation{NTT Basic Research Laboratories, NTT Corporation, 3-1 Morinosato-Wakamiya, Atsugi, Kanagawa, 243-0198, Japan}
\affiliation{National Institute of Informatics, 2-1-2 Hitotsubashi, Chiyoda-Ku, Tokyo 101-8430, Japan}
\author{Robert B. Mann}
\email{rbmann@uwaterloo.ca}
\affiliation{Department of Physics \& Astronomy, University of Waterloo,  Ontario Canada N2L 3G1}

\begin{abstract}
An interferometric experiment is described that characterizes an optical cat state in a cavity mode. Our method describes how to measure the amplitude and phase of the different coherent states that make up the cat states. We show that using a recently proposed quantum nondemolition measurement technique, we can circumvent the problem of decoherence inherent in measurement involving superposition of coherent states. 
\end{abstract}

\maketitle

\section{Introduction}
The last century saw the discovery of  quantum mechanics, a set of  principles that describes nature  at the microscopic level. These principles of superposition and entanglement have had a profound effect on our understanding of our natural world and of realism, and are now leading to a technological revolution. However many fundamental and foundational questions still remain, with one of the most profound being the quantum/classical boundary (if one actually exists). 

At the heart of this issue are superposition states of macroscopic objects. The process of observing superposition in macroscopic systems \cite{S1935} has attracted a great deal of attention as these superpositions exhibit important interference effects which will be very useful in quantum information processing \cite{gkw,ampdamp,Vlastakis607} and quantum metrology \cite{weakforce,laghaout}. For a long time, observing quantum superpositions was considered almost impossible because of decoherence.  Large systems are in constant interaction with their environment; they form a larger ensemble with the environment and, as such, lose their quantum nature.

Considerable technological progress has been made in realizing an ideal system in which we can study the decoherence effect in a well-controlled environment \cite{cavities1,prl57,opsch,pra1992}. One way of detecting macroscopic quantum superposition and successfully observing its decoherence, is to trap the state in a high-quality cavity and probe it with a detector   with a long decay lifetime \cite{cavities1}.  Mesoscopic fields stored in a high-quality superconducting cavities in the microwave region  are ideal tool to reveal the decoherence process. Their interaction with a single circular Rydberg atom prepares them in a superposition of fields, containing a few photons, with different classical phases. One can later probe the evolution of the resultant Schroedinger cat state with another atom \cite{ HWG,Haroche,Davidovich,cavities1}.

The simplest macroscopic quantum superposition that has been widely prepared and vastly  studied is a superposition of two coherent states \cite{coinco}. We shall describe here an experimental proposal that detects a general superposition of two coherent states and its subsequent decoherence  into a statistical mixture.  The scheme is based on mode invisibility measurement, a quantum non-demolition (QND) measurement \cite{secondemo} idea that maximizes the geometrical structure of the quantum field stored in the cavity mode \cite{marvy2013,marvy2014}.
\begin{figure}[b]
\includegraphics[width=.45\textwidth]{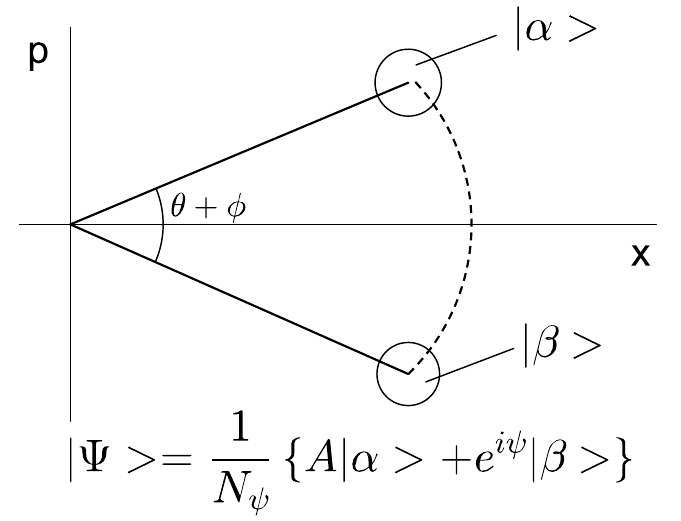}
\caption{Phase-space representation of two coherent states of identical mean photon number 
$|\alpha|=|\beta|$ but different phases. } 
\label{fig1}
\end{figure}
\twocolumngrid

\section{Schroedinger Cat State }\label{one}

A coherent state $\ket{\alpha}$ is the eigenstate of the annihilation operator $\hat{a}$ with eigenvalue $\alpha$: $\hat{a}\ket{\alpha} = \alpha \ket{\alpha}$, 
\begin{align}\label{rho}
 \ket{\Psi_{ \psi}} = \frac{1}{N_{\psi}}\Big(A\ket{\alpha} + e^{\ii \psi} B\ket{\beta}\Big),
\end{align}
where $\alpha = |\alpha|e^{\ii \theta}$ is a complex amplitude with magnitude $|\alpha|$ and phase $\theta$.  We define a general cat state to be a superposition of two coherent states  $\ket{\alpha}$ and $\ket{\beta}$ respectively with size $|\alpha-\beta|^{2}$, where $\alpha $ and $\beta $ are the coherent amplitudes, and  the phase $\psi$ reveals the coherence of the state superposition (see Fig.~\ref{fig1}). In this definition, the real factors $A$ and $B$ are such that $A^2+B^2=1$, while the normalization factor 
\begin{align*}
N_{\psi}=  (1+ 2 A B\operatorname{Re}[ \langle \beta \ket{\alpha}e^{-\ii\psi}])^{1/2}
\end{align*}
measures the nonorthogonality of the two coherent states with overlap $\langle \beta | \alpha \rangle = e^{-1/2(|\alpha|^{2} + |\beta|^{2}-2\alpha \beta^{*})}$. In the density operator formalism, this state is defined as
\begin{align}\label{ss}\nonumber
\rho = \frac{1}{N_{\psi}^{2}}&\Big(A^{2}\ket{\alpha}\langle \alpha| + B^{2}\ket{\beta}\langle \beta| +e^{\ii \psi} A B\ket{\beta}\langle \alpha| \\ 
&+ e^{-\ii \psi} A B\ket{\alpha}\langle \beta| \Big)
\end{align}
A statistical mixture of the coherent states is given as
\begin{align}\label{sm}
\rho_{\text{sm}} = \frac{1}{N^{2}} \Big( A^{2}\ket{\alpha}\langle \alpha| + B^{2}\ket{\beta}\langle \beta|  \Big)
\end{align}
Comparing equations \eqref{ss} and \eqref{sm} respectively, we see the phase factor $\psi$ appears due to the quantum interference between the two coherent states \cite{coherentstate}.  This captures the quantum nature of the superposition of coherent states \eqref{rho}.  The properties of the state \eqref{rho} have not been studied except for special cases when $\psi = 0, \pi$ and $\pi/2$. In these special cases, we define three different cat states CS :
\begin{subequations}\label{fours}
\begin{align}
 \ket{\Phi_{0}} &= N_{0}^{-1}\Big(A\ket{\alpha} +  B\ket{\beta}\Big) \label{ev-s}\\
 \ket{\Phi_{\pi }} &= N_{\pi}^{-1}\Big(A\ket{\alpha} -  B\ket{\beta}\Big) \label{od-s}\\
 \ket{\Phi_{\pm\frac{\pi}{2}}} &= N_{\pm \pi/2}^{-1}\Big(A\ket{\alpha}  \pm \ii B \ket{\beta}\Big) \label{ys-s}
\end{align}
\end{subequations}
If we set $\beta = -\alpha$, these states yield the even, odd, and Yuker-Stoler CS respectively \cite{prl57,buzek,schleich}. The statistical properties of these three states have been extensively studied \cite{buzek,schleich}. The even and odd CS are orthogonal to one another which suggest that it is possible to distinguish between the two states. In the Fock basis the even and odd CS are each represented as a superposition of only even and odd number states respectively, which means that the photon-number distribution of the two states exhibit significant oscillations \cite{buzek} so that it is possible to distinguish the two states through a photon counting measurement and by resolving the oscillations. As  is generally known, in a photon counting measurement, photons get destroyed in the process  thereby significantly destroying the quantum system at hand. We will describe in this paper a possible non-destructive measurement that distinguishes between the even and odd cat states. The Yuker-Stoler cat state is slightly different from the even and odd cat states but also exhibits the same interesting and useful features.

A more general class of quantum states of light is obtained by applying squeezing operations \cite{squeezedlight}.  
Applying the single mode squeeze operator 
$$
\hat{S}(\zeta) = \exp\Big(\frac{1}{2}(\zeta^{*}\hat{a}^{2} - \zeta \hat{(a^{\dagger})^{2}})\Big)
$$ 
on the cat states results in
\begin{align}\label{sqstate}
\ket{\Psi_{\psi}} = \frac{1}{N_{\psi}}\Big(AS (\zeta)\ket{\alpha} +e^{\ii \psi} B S(\zeta)\ket{\beta}\Big),
\end{align}
where  $\zeta = r e^{\ii \delta}$ is the  complex squeezing parameter, $r$ the amplitude and $\delta$ the phase.

Having discussed briefly the quantum superposition of coherent states and their features, we will proceed in the next section in describing how to detect the states assuming we have the system in a cavity. 
\begin{figure}[htb]
\includegraphics[width=.45\textwidth]{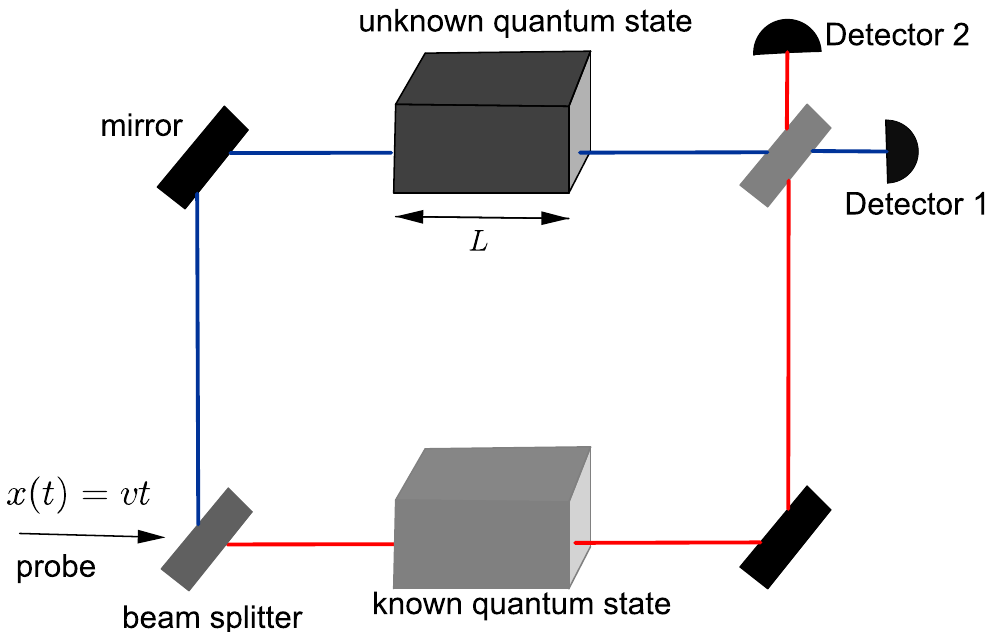} 
\caption{(Color online): Measurement setup for the recently proposed mode-invisibility measurement technique \cite{marvy2013} } 
\label{fone}
\end{figure}

\subsection{Detecting the Cat State}

Our goal in this article is to see how one can detect the cat states non-destructively and distinguish between the different types with reliable probability.  To this end we apply the recently proposed measurement technique-- Mode-Invisibility Measurement Scheme  \cite{marvy2013}. The measurement setup, which is an atomic interferometer, is   shown in Fig \ref{fone}. Two cavities, each of length   $L$, are placed along the branches of the interferometer. We populate a single mode $n$ of   one cavity with an unknown state of light $\ket{\Psi_{n}}$ and the other with a known state of light $\ket{\Psi_{n}^{R}}$. We assume that the remaining modes are empty (i.e. in   the vacuum state). A probe in the form of a two level atom initially in its ground state is sent  at speed $v$ through one input mode of a beam splitter (BS), whose action is to split the initial state configuration $\ket{g,0}$  into the superposition   $\frac{1}{\sqrt{2}}\ket{g,0}_{1}$ and $ \frac{1}{\sqrt{2}}\ket{0,g}_{2}$  (labeling the interferometric paths $ i = 1, 2$) that  travel along the interferometric branches and encounter one of the two cavities along the way.
 
In the mode-invisibility scheme, it is assumed that the probe interacts  adiabatically with only the even cavity mode for a very short time. The cavity state is not significantly perturbed  but the probe acquires a global phase factor $\gamma_{i}$, which carries path-dependent information about the quantum state in the cavity. After the interaction time $T=L/v$,  the probe exits the cavity and recombines at the second BS. The probability of detecting each partial probe atom is given as
\begin{align}
P_{\pm} = \frac{1}{2} \Big(1 \pm \cos[\Delta \gamma]\Big)
\end{align}
where $\Delta \gamma = \gamma_{1} - \gamma_{2}$ is the phase difference acquired along each interferometric path. The subsequent measurement of $\Delta \gamma$  thus yields information about the unknown quantum state.  

The mode-invisibility measurement scheme takes advantage of  ideas underlying the development of quantum non-demolition (QND) measurement \cite{marvy2013}. Assuming  the two-level   probe is initially in its ground state, we require that the probe remains in its ground state (up to a phase) after its path-dependent interaction with the cavity state, commensurate with a QND measurement scheme.  This implies that the probability  the probe gets excited after interaction remains approximately zero 
\begin{align}\label{tr}
P_{\ket{e}} = \langle e| \textrm{Tr}_{f}[U_{i}\rho U_{i}^{\dagger}]|e\rangle \ll 1
\end{align}
along either path.  Hence the final state of the system is very approximately equal to the initial state 
\begin{align}\label{two}
\ket{\Psi(T)} = U_{i}(T,0)\ket{\Psi(0)} \approx e^{\ii \gamma_{i}}\ket{\Psi(0)}
\end{align} 
except for a global path-dependent dynamical phase given as $\gamma_{i}$.
From equation \eqref{two}, one can obtain $\gamma_{i}$ to be:  
\begin{align}\label{gammar}
\gamma_{i} = \operatorname{Re}[- \ii \log\lbrace \langle \Psi(0)|U_{i}(T,0)|\Psi(0)\rangle\rbrace]
\end{align}
This phase $\gamma$ is  a global phase which is not measureable in experiment. The atomic interferometer described above can help us obtain a measurement for $\gamma$ when we define a reference quantum state in the second cavity. For simplicity, we use the vacuum state as a reference state.  As we will show, the interference between the component states is a function of the relative phase difference $\gamma$ and can generate a number of novel effects.

\section{Measurement Procedure}

Suppose we have a superposition of coherent states \cite{coinco} $\ket{\alpha_{n}} $ and $\ket{\beta_{n}}$ prepared in an even cavity mode $n$ with frequency $\omega_{n} = n \pi c/ L$ with $n=1,2,\cdots$.  The initial  state of our system is given by the product state of the probe with the cavity field:
\begin{align}\label{qstate}
\ket{\Psi_{\psi}}(0) = \ket{g} \otimes  \frac{1}{N_{\psi}}\Big(AS_{n}(\zeta)\ket{\alpha_{n}} +e^{\ii \psi} B S_{n}(\zeta)\ket{\beta_{n}}\Big).
\end{align}
where $\ket{g}$ is the ground state of the probe and we have applied the squeezing operator $S_{n}(\zeta)$ to the superposition of coherent states. We model the probe-field interaction with the known Unruh-Dewitt Hamiltonian  $\hat{H}_{\text{int}}$ \cite{unruh}:
\begin{align*}
\hat{H}_{\text{int}} =   \lambda \hat{\mu}(t) \hat{\phi}[x(t)]\\
\end{align*}
where $\lambda$ is the coupling constant,  $\phi[x(t)]$ is  field to which the detector is coupling
(which for simplicity we take to be a massless scalar), and $\mu(t)$ is the monopole moment of the detector, which has the usual form 
\begin{align*}
\hat{\mu}(t) = (\sigma^{+} e^{\ii \Omega t} + \hat{\sigma}^{-} e^{- \ii \Omega t})
\end{align*}
in the interaction picture, where $\Omega$ is the probe's transition frequency and $\sigma^{+} (\sigma^{-})$ its raising (lowering) operator.   Expanding the cavity mode in the Dirichlet boundary condition yields the Hamiltonian in the interaction picture
\begin{align*}
\hat{H}_{\text{int}} = \sum_{j=1}^{\infty}  &  \frac{\lambda \hat{\mu}(t)}{\sqrt{k_{j} L}} \Big( a^{\dagger}_{j}e^{\ii \omega_{j} t} + a_{j} e^{-\ii \omega_{j} t} \Big)\sin[k_{j}x(t)]
\end{align*}
where $a^{\dagger}(a)$ is the creation (annihilation) operator for the field state.  The atom-field evolution in the cavity from the interaction time $t=0$ to time $t=T$ is governed by the evolution operator
\begin{align}\label{unitary}
&U(0,T) =  \mathcal{T}\text{exp}\bigg[-\ii\int_{0}^{T} dt\hat{ H}_{\text{int}}(t)\bigg]  \\
&\quad = I \underbrace{- \ii \int_{0}^{T}dt\hat{ H}_{\text{int}}(t)}_{U^{(1)}}  \underbrace{-\int_{0}^{T} dt \int_{0}^{t}d\tau  H_{int}(t) H_{int}(\tau)}_{U^{(2)}}+ \cdots
\nonumber
\end{align}
and so the system's evolution operator at time $T$ is 
\begin{align*}
\rho_{T} = [1 + \hat{U}^{(1)}+\hat{U}^{(2)}  ]\rho_{0}  [1 + \hat{U}^{(1)}+\hat{U}^{(2)}  ]^{\dagger} + O(\lambda^{3})
\end{align*}
where $\rho_{0}$ is the density operator for the state \eqref{qstate}. 

\subsection{Transition Probability}

 To check that the probe remains in its ground state after the interaction time $T$, we will compute its excitation transition probability which is given up to second order in $\lambda$, by the expression
\begin{align}\label{transprob}\nonumber
P_{\ket{e}}& = \langle e |\operatorname{Tr}_{F}[  \rho(0) + U^{(1)}\rho + \rho U^{(1)\dagger}+ U^{(2)}\rho + \rho U^{(2)\dagger}]\ket{e}\\
&+  \langle e |\operatorname{Tr}_{F}[  U^{(1)} \rho(0) U^{(1) \dagger}   ]\ket{e}
\end{align}
where 
\begin{align*}
\nonumber U^{(1)}  &
= \frac{ \lambda}{\ii} \sum_{j}( \sigma^{+} a_{j}^{\dagger}I_{+,j} + \sigma^{+} a_{j}I_{-,j}^{*})\\
 U^{(2)}  
=&- \lambda^{2} \sum_{j}\sum_{l}\bigg( \sigma^{-}\sigma^{+} a_{j}^{\dagger}a_{l}^{\dagger}I_{-,j} \circ I_{+,l} +\sigma^{-}\sigma^{+} a_{j}^{\dagger}a_{l}I_{-,j}\circ I_{-,l}^{*}\\
&+  \sigma^{-}\sigma^{+} a_{j}a_{l}I_{+,j}^{*}\circ I_{-,l}^{*} + \sigma^{-} \sigma^{+} a_{j}a^{\dagger}_{l}I_{+,j}^{*}\circ I_{+,l}\bigg)
\end{align*}
follow from expanding the time evolution operator \eqref{unitary}, and we have neglected the terms that do not contribute to the equation \eqref{transprob}.   For notational convenience we have defined
\begin{align}\label{otu}
I_{\pm, j}&=  \frac{1}{\sqrt{k_{j}L}}\int_{0}^{T} ~\text{dt}~ e^{\ii( \omega_{j} \pm \Omega)t}\sin[k_{j}x(t)]\\\label{abuo}
I_{\pm, j}\circ I_{\pm,l}&=  \frac{1}{\sqrt{(k_{j}L)(k_{j}L)}}\int_{0}^{T} ~\text{dt}~   \int_{0}^{t} ~\text{d}\tau~e^{\ii( \omega_{j} \pm \Omega)t}e^{\ii( \omega_{l} \pm \Omega)\tau}\\& \times \sin[k_{j}x(t)]\sin[k_{l}x(\tau)] \nonumber
\end{align}
with the latter relation defining the $\circ$ operation.
 
The ground $\ket{g}$ and excited $\ket{e}$ states of the probe  are orthogonal states i.e. $\langle e\ket{g} = 0$. Given the initial state \eqref{qstate}, the terms in the first line of \eqref{transprob} vanish. Therefore the leading order contribution to the probe's excitation transition probability is 
\begin{align}\label{transprob2}
P_{\ket{e}} = \langle e |\operatorname{Tr}[ U^{(1)} \rho(0) U^{(1) \dagger}]\ket{e}
\end{align}
which upon evaluation, yields the expression
\begin{widetext}
\begin{align}\label{transitionp}\nonumber
P_{\ket{e}}& = {\lambda^{2}} \Bigg \lbrace  \sum_{ m} |I_{+, m}|^{2} + \frac{ |I_{+,n}|^{2}}{N^{2}_{\psi}} \Big(  \textsf{S}_{r}^{2} +(\textsf{C}_{r}^{2} + \textsf{S}_{r}^{2}) (A^{2}|\alpha|^{2} + B^{2}|\beta|^{2})  - 2 \textsf{C}_{r} \textsf{S}_{r} ( A^{2} \operatorname{Re}\lbrace e^{-\ii \delta} \alpha^{2} \rbrace  +  B^{2}  \operatorname{Re}\lbrace e^{-\ii \delta} \beta^{2} \rbrace )\\\nonumber
&\qquad\qquad \qquad  + 2 A B [S^{2}_{r} \operatorname{Re}[e^{-\ii \psi}\langle \beta_{n}\ket{\alpha_{n}}] +  (\textsf{C}_{r}^{2} + \textsf{S}_{r}^{2}) \operatorname{Re}[e^{-\ii \psi}(\beta^{*}\alpha)\langle \beta_{n}\ket{\alpha_{n}}]\\
&\qquad \qquad \qquad\qquad -   \operatorname{Re}\lbrace e^{-\ii \delta}e^{\ii \psi} \beta^{2}\langle \alpha_{n}\ket{\beta_{n}} \rbrace \textsf{C}_{r} \textsf{S}_{r}-   \operatorname{Re}\lbrace  e^{-\ii \delta} e^{-\ii \psi} \alpha^{2}\langle \beta_{n}\ket{\alpha_{n}} \rbrace \textsf{C}_{r} \textsf{S}_{r}]  \Big)  \Bigg \rbrace ,
\end{align}
\end{widetext}
where $\textsf{C}_r = \cosh r$ and $\textsf{S}_r = \sinh r$. The first term corresponds to vacuum fluctuations in the cavity modes. The second term corresponds to  typical counter-rotating contributions $(\Omega + \omega_{j})$ that describe an interaction process where the probe gets excited and a single photon is emitted in the cavity mode $j$. 

In evaluating \eqref{transitionp}, we applied the mode invisibility technique \cite{marvy2013} to eliminate the rotating-wave contribution $(\omega_{j} - \Omega)$, which describes the physical process where the probe absorbs a photon from the field mode $j$. To see how this works, we note that  the integral $I_{\pm, j}$  is easily solvable and it gives \cite{marvy2013}
\begin{align*}
I_{\pm j} = \frac{\left[e^{\ii\frac{L}{v}(\omega_j \pm \Omega )}(-1)^{j} -1\right] L v\sqrt{j \, \pi} }{\left(j \, \pi v\right)^2-L^2(\omega_j \pm\Omega)^2}.
\end{align*}
We see from this expression, that when we allow a resonant interaction $(\omega_{j} = \Omega)$ and set the detector to probe only the even cavity mode $j=2n$, then   the rotating wave term $I_{-, j} $ vanishes but not the counter-rotating wave term $I_{+, j} $. We can improve this result by controlling the speed at which the probe crosses the cavity. If we chose $j=2n$ to be an even number and the atomic speed to be approximately $v=j c/N$ for $N=1,2,3 \cdots$, we can cancel the contribution of the counter-rotating wave term to the transition probability, thereby eliminating any dependence it has on the probe mode. This makes the cavity mode completely invisible to the probe and so we have the contributions of the vacuum terms in equation \eqref{transitionp}. For realistic physical parameters $v=1000$ m/s, $\lambda = 10^{11} , \Omega = 10^{-4}\lambda$, we find $P_{\ket{e}} \approx 10^{-21}$. We are thus able to control the excitation of the probe whilst preserving the largest contribution to its phase after the interaction.
\bigskip

\subsection{Estimating the atomic phase by the mode-invisivibility method}

 Based on the discussion in the previous section, we can assume that the measurement condition \eqref{tr} is satisfied and our quantum system evolves according to relation \eqref{two}. We now proceed to compute $\Delta \gamma$. According to equation \eqref{gammar}, we computed the phase $\gamma_{1}=\operatorname{Re}[\eta_{1}]$ that the probe acquires on interacting with the general squeezed cat state in cavity mode. This is given by
\begin{widetext} 
\begin{align}\label{result1}\nonumber
\eta_{1}(\psi)& = - \ii \text{ln} \Big \lbrace 1 - \lambda^{2} \sum_{m}I_{+,m}^{*}\circ I_{+,m} -\frac{ \lambda^{2}}{N^{2}} I_{+,n}^{*}\circ I_{+,n}\Big(  \textsf{S}_{r}^{2} +(\textsf{C}_{r}^{2} + \textsf{S}_{r}^{2}) (A^{2}|\alpha|^{2} + B^{2}|\beta|^{2})  - 2 \textsf{C}_{r} \textsf{S}_{r} ( A^{2} \operatorname{Re}\lbrace e^{-\ii \delta} \alpha^{2} \rbrace  \\\nonumber
&+  B^{2}  \operatorname{Re}\lbrace e^{-\ii \delta} \beta^{2} \rbrace )+ 2 A B \Big[\textsf{S}_{r}^{2} \operatorname{Re}[e^{-\ii \psi}\langle \beta_{n}\ket{\alpha_{n}}] +  (\textsf{C}_{r}^{2} + \textsf{S}_{r}^{2}) \operatorname{Re}[e^{-\ii \psi}(\beta^{*}\alpha)\langle \beta_{n}\ket{\alpha_{n}}] -   \operatorname{Re}\lbrace e^{-\ii \delta}e^{\ii \psi} \beta^{2}\langle \alpha_{n}\ket{\beta_{n}} \rbrace \textsf{C}_{r} \textsf{S}_{r} \\
&-   \operatorname{Re}\lbrace  e^{-\ii \delta} e^{-\ii \psi} \alpha^{2}\langle \beta_{n}\ket{\alpha_{n}} \rbrace \textsf{C}_{r} \textsf{S}_{r}\Big]  \Big) \Big \rbrace
\end{align}
\end{widetext}
 where   $I_{+,m}^{*}\circ I_{+,m} $ is defined analogously to  \eqref{abuo}.  We see from equation \eqref{result1} that there is a small interference term taking non-zero values, but it is of the order of the scalar product of the two cat components $\ket{\alpha}$ and $\ket{\beta}$, which  is vanishingly small as soon as they are separated.  This implies that our method is only able to detect the interference between the two cat components for small cat size. However our ability to distinguish between these states is advanced when we consider squeezing.
 
$\eta_{1}(\psi)$ in general is a global phase shift which we cannot detect. However with the measurement setup atomic interferometer in Fig \ref{fone} where a reference state to be placed in the cavity along the other path \cite{Gravwaves}. For simplicity, we assume this to be the vacuum state $\ket{0_{R}}$. The phase $\gamma_{2}$ acquired along this second path is
\begin{align}\label{sec}
\eta_{2} = - \ii \operatorname{ln}\Big[1-\lambda^{2} \sum_{m}I_{m}\circ I_{+,m}\Big]
\end{align}
yielding  
\begin{align}\label{fin}
\Delta \gamma &= \operatorname{Re}[\eta_{1}(\psi)-\eta_{2}]
\end{align}
for the interferometric phase difference. 

To maximize $ \Delta \gamma$, we assumed that the probe's frequency is on resonance with the populated cavity mode.  In general this resonant interaction leads to a violation of the criteria \eqref{tr}.  However if the probe interacts only with the even cavity mode eq. \eqref{tr} can be satisfied.  Recall that allowing the probe to interact with only the even cavity modes eliminates the leading contribution to the excitation transition probability. This is the idea behind the mode-invisibility measurement technique -- the effect of the probe-cavity mode interaction does not significantly alter the state of the cavity mode even while information about the cavity mode is registered on the probe's state.  In the next section, we will discuss how we can use this interferometric phase difference to distinguish between the even, odd, and Yuker-Stoler states.

\section{Distinguishing The Different Cat States}
In the previous section, we have derived general expressions for the interferometric phase difference of a pair of superpositions of squeezed coherent states \eqref{result1}.To clarify our analysis, in what follows we study the properties associated with an interaction between the two-level probe and a superposition of two coherent states $\ket{\alpha}$ and $\ket{-\alpha}$ with squeezing and non-squeezing respectively. If  we set values of $A = B = 1/\sqrt{2}$ and $|\beta| = |\alpha|,~ \phi = -\theta$ and $\theta = \pi/2$ in equation \eqref{qstate}, the initial state is of the form:
\begin{align}\label{statesim}
\ket{\Psi_{\psi}} =&   \lbrace1+ \cos(\psi) e^{-2|\alpha|^{2}} \rbrace^{-1/2}\ket{g}\otimes S(r)[\ket{\alpha_{\kappa}} + e^{\ii \psi} \ket{-\alpha_{\kappa}}]
\end{align}
With the probe in its ground state $\ket{g}$ interacting with the cat state (CS) trapped in a cavity mode $\kappa$, we have shown in \eqref{result1} that the probe is able to acquire a phase factor provided that the excitation transition probability is approximately zero. We will discuss the behaviour of this phase factor for the different cat states respectively. 

The phase difference acquired by a detector that interacts with a cavity sustaining a squeezed cat state \eqref{statesim} can be obtained from \eqref{result1} and is given as 
\begin{widetext} 
\begin{align}\label{result}\nonumber
\eta_{1}(\psi,r)& = - \ii \text{ln} \Big \lbrace 1 - \lambda^{2} \sum_{m}I_{+,m}^{*}\circ I_{+,m} -\frac{ \lambda^{2}}{N^{2}} I_{+,n}^{*}\circ I_{+,n}\Big[  \textsf{S}_{r}^{2} +(\textsf{C}_{r}^{2} + \textsf{S}_{r}^{2}) |\alpha|^{2}(1+e^{-2|\alpha|^{2}\sin^{2}\theta}\cos(\psi - 2\theta-|\alpha|^{2}\sin 2\theta)) \\ &- 2 \textsf{C}_{r} \textsf{S}_{r}|\alpha|^{2}  \cos (\delta)\Big(\cos(2\theta) +e^{-2|\alpha|^{2}\sin^{2}\theta}\cos(-\psi + 2\theta+|\alpha|^{2}\sin 2\theta) \Big) + \textsf{S}_{r}^{2} e^{-2|\alpha|^{2}\sin^{2} \theta}\cos(\psi -|\alpha|^{2}\sin 2\theta)  \Big] \Big \rbrace
\end{align}
\end{widetext}

 The interference term $\mathrm{e^{-2|\alpha|^{2}\sin^{2}\theta}\cos(\psi - 2 \theta - |\alpha|^{2}\sin(2\theta))}$ in \eqref{result} arises because we have a superposition  \eqref{statesim} of two coherent states.  Substituting equations \eqref{result} and \eqref{sec} in \eqref{fin}, we  obtain the interferometric phase difference $\Delta \gamma$, which reveals the quantum nature of the superposition of two coherent states upon detection. We can enhance the observability of the interference term by adjusting the phase $\theta$ of the coherent states that make up the cat states such that $\sin(\theta) = 0$.  From the exponential $e^{-2|\alpha|^{2}\sin^{2}\theta}$, we see that the interference term rapidly fades as $\sin^{2}(\theta)$ becomes larger than $1/2|\alpha|^{2}$.  Hence an experimentalist can verify through the mode-invisibility technique that the state has macroscopically distinguishable components without significantly destroying the states. 

 \subsection{Relationship between $\Delta \gamma$ vs $|\alpha|$ for the different cat states.}\label{alpha}
The intensity of light in an even, odd and Yuker-Stoler state is given respectively as
\begin{subequations}\label{eqone}
\begin{align}
\overline{n}_{\text{ev}} = &|\alpha|^{2}\Bigg( \frac{ 1- e^{-2|\alpha|^{2}}}{1+ e^{-2|\alpha|^{2}}}\Bigg)\\
 \overline{n}_{\text{od}} =& |\alpha|^{2}\Bigg( \frac{ 1+ e^{-2|\alpha|^{2}}}{1- e^{-2|\alpha|^{2}}}\Bigg) \\
 \overline{n}_{\text{YS}} =& |\alpha|^{2}
\end{align}
\end{subequations}
 Suppose the coherent state that makes up the CS has phase $\theta = \pi/2$. For the even, odd and Yuker-Stoler CS with $\psi=0,\pi$ and $\psi=\pm \pi/2$ respectively, equation \eqref{result} yields
\begin{subequations}\label{sqphase}
\begin{align}\nonumber
\eta_{1}(0)& = - \ii \text{ln} \Big \lbrace 1 - \lambda^{2} \sum_{m}I_{+,m}^{*}\circ I_{+,m} -\lambda^{2} I_{+,n}^{*}\circ I_{+,n}\\& \times \Big[  \textsf{S}_{r}^{2} +(\textsf{C}_{r}^{2} + \textsf{S}_{r}^{2}) \overline{n}_{\text{ev}} + 2 \textsf{C}_{r} \textsf{S}_{r}  |\alpha|^{2}\cos (\delta) \Big] \Big \rbrace,\\\nonumber
\eta_{1}(\pi)& = - \ii \text{ln} \Big \lbrace 1 - \lambda^{2} \sum_{m}I_{+,m}^{*}\circ I_{+,m} -\lambda^{2} I_{+,n}^{*}\circ I_{+,n}\\& \times \Big[  \textsf{S}_{r}^{2} +(\textsf{C}_{r}^{2} + \textsf{S}_{r}^{2}) \overline{n}_{\text{od}}+ 2 \textsf{C}_{r} \textsf{S}_{r} |\alpha|^{2} \cos (\delta) \Big] \Big \rbrace,\\\nonumber
\eta_{1}(\frac{\pm \pi}{2})& = - \ii \text{ln} \Big \lbrace 1 - \lambda^{2} \sum_{m}I_{+,m}^{*}\circ I_{+,m} -\lambda^{2} I_{+,n}^{*}\circ I_{+,n}\\
&\times \Big[  \textsf{S}_{r}^{2} +(\textsf{C}_{r}^{2} + \textsf{S}_{r}^{2}) |\alpha|^{2} + 2 \textsf{C}_{r} \textsf{S}_{r}  \cos (\delta)|\alpha|^{2} \Big] \Big \rbrace,
\end{align}
\end{subequations}
We see the equations depending on the intensities of light in the different CS.
Substituting equations \eqref{sqphase} and \eqref{sec} in \eqref{fin}, we   therefore see that the various cat states that can be formed from superposition of two coherent states $\ket{\alpha}$ and $\ket{-\alpha}$ have distinct interferometric phase differences $\Delta \gamma$ which depend on the intensities of light in the different cat states. That $\Delta \gamma$ obtainable from each cat state depends on the intensities of light in the state marks an important signature that experimentalists can measure  and use to describe the characteristic features of these cat states.

\begin{figure}[h!]
 \begin{minipage}{.5\textwidth}  \includegraphics[width=.5\textwidth]{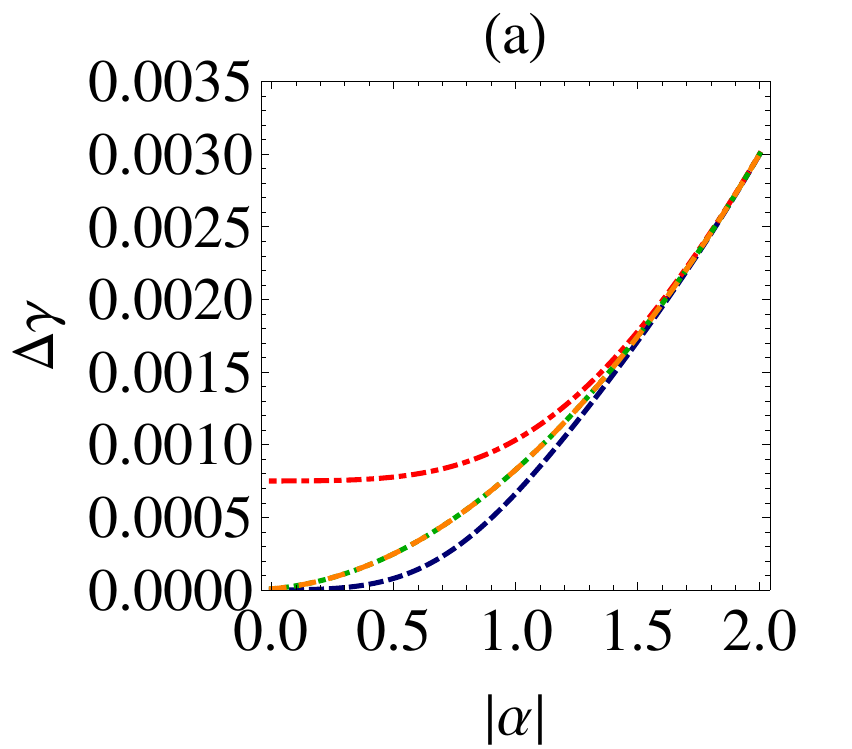} \includegraphics[width=.48\textwidth]{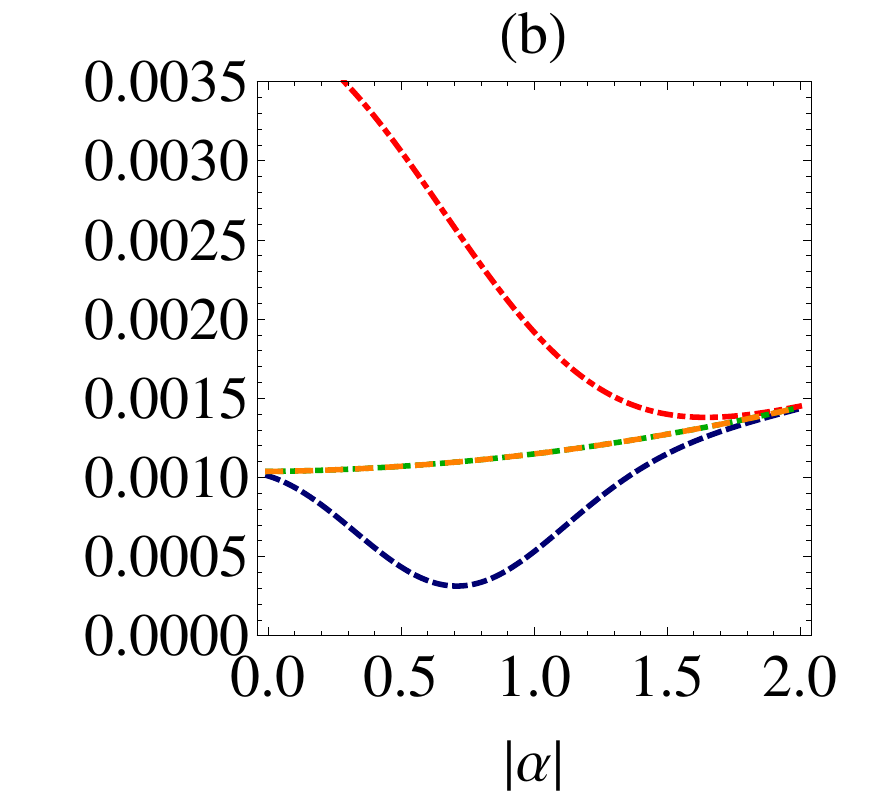}
  \caption{(Color online): Relative phase difference $\Delta\gamma$ vs the parameter $|\alpha|$ for the different cat states: even  (blue dashed line), odd  (red dotdashed line),  and Yuker-Stoler $\pm \pi/2$ (green dotted and orange medium dashed lines)  respectively. Figure (a) shows this relationship in the absence of squeezing $\zeta=0$ and  figure (b) shows the relationship for small squeezing $\zeta =|1|$. The coupling constant $\lambda = 10^{-4}\Omega \text{Hz}$ where the atomic transition frequency of the probe $\Omega =  10^{11}\text{Hz}$. Here we consider the phase $\theta = \pi/2$ for a maximum output of our result. }
\label{onep}
 \end{minipage}
\end{figure}
 We illustrate in Fig.~\ref{onep} the dependence of $\Delta \gamma$ vs $|\alpha|$ for  squeezing parameters $\zeta = 0 $ and $\zeta= |1|$ respectively.  We see that for sufficiently small values of $|\alpha|$, we can distinguish between the even, odd and Yuker-Stoler cat states respectively. It is obvious that squeezing affords better ability to distinguish between the three cat states as the curves are well separated from one another. The differences among the even, odd and Yuker-Stoler cat states decreases with increasing $|\alpha|$ and completely disappears even at $|\alpha| \approx 2$ (see Fig. \ref{onep}).  Of course this is true for certain values of $\theta$, the relative phase of the coherent state.  We see that we are not able to distinguish between the Yuker-Stoler $\pm \pi/2$ states when the phase $\theta = n\pi/2$, where $n =1,2,3 \cdots$. However this changes for values of $\theta=n\pi$ as we are able to distinguish these two states. 
 
\subsection{Relationship between $\Delta \gamma$ vs the squeezing parameters $r,\delta$ }
In section \ref{alpha}, we discussed how the even, odd and Yuker-Stoler CS could be distinguished by looking at the relationship between the interferometric phase difference $\Delta \gamma$ and the parameter $|\alpha|$. We considered two cases when the squeeze parameter $|\zeta|=0$ and $1$ respectively.  Here we want to see to what extent the states could be distinguished with respect to the squeezing parameters $r$ and $\delta$ respectively
\begin{figure}[h!]
 \begin{minipage}{.5\textwidth}  \includegraphics[width=.5\textwidth]{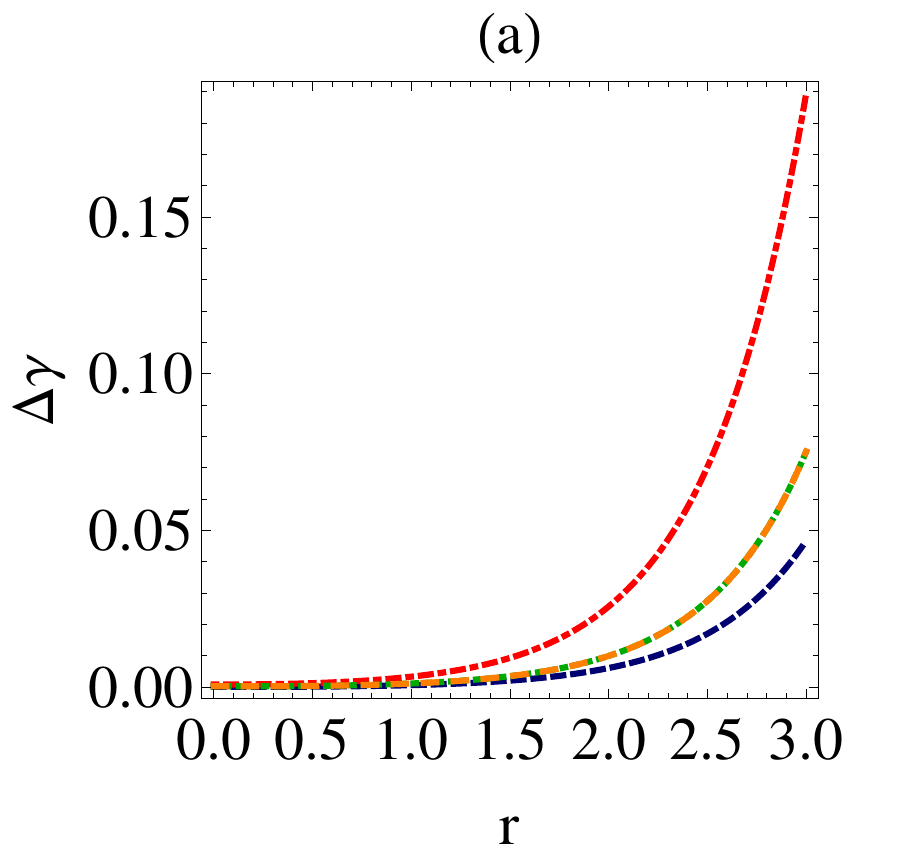} \includegraphics[width=.48\textwidth]{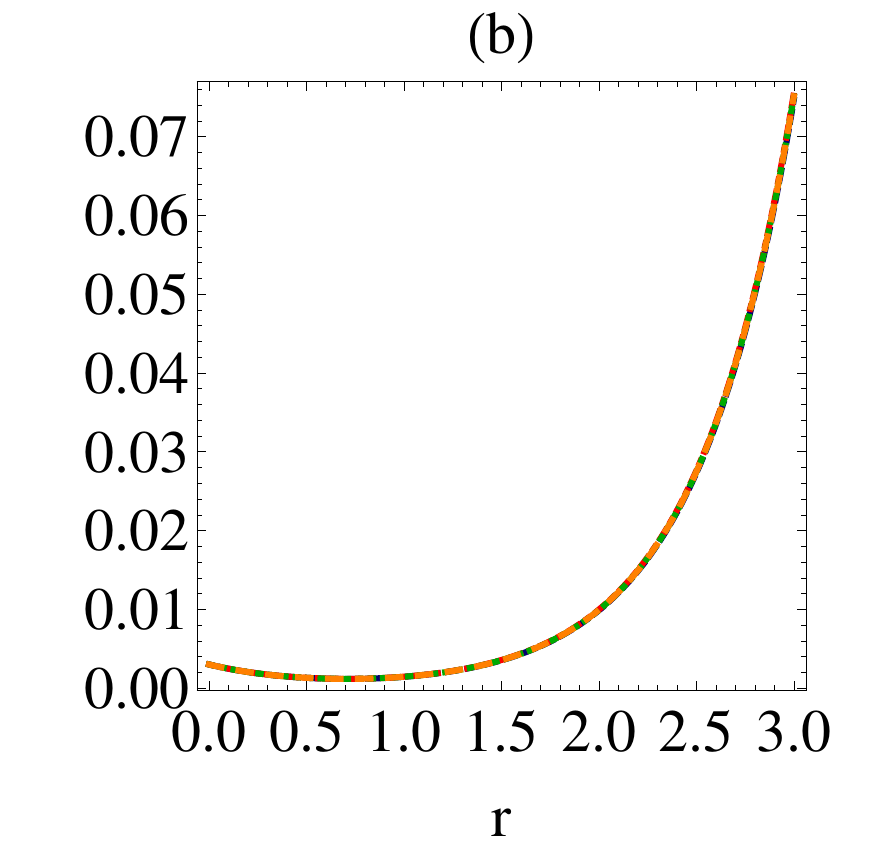}
  \caption{(Color online):  $\Delta\gamma$ vs the parameter $r$ for the different CS: even  (blue dashed line), odd  (red dotdashed line),  and Yuker-Stoler $\pm \pi/2$ (green dotted and orange medium dashed lines)  respectively. In Fig (a), $|\alpha|=0.5$, (b) $ |\alpha| = 2$ and $\delta = \pi$ }
\label{oner}
 \end{minipage}
\end{figure}

\begin{figure}[h!]
 \begin{minipage}{.5\textwidth}  \includegraphics[width=.5\textwidth]{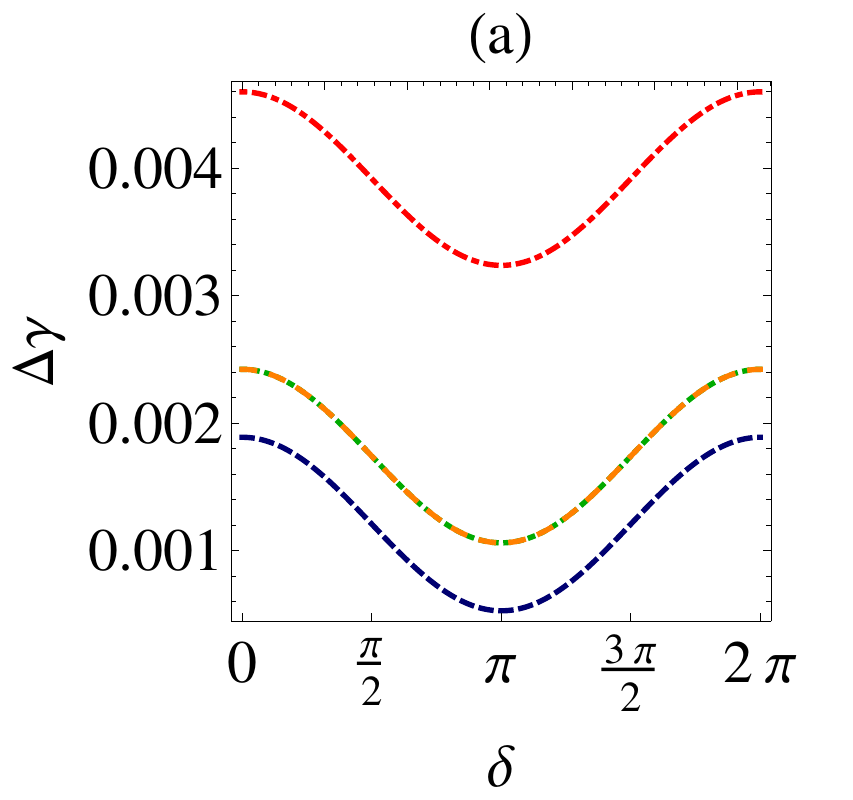} \includegraphics[width=.48\textwidth]{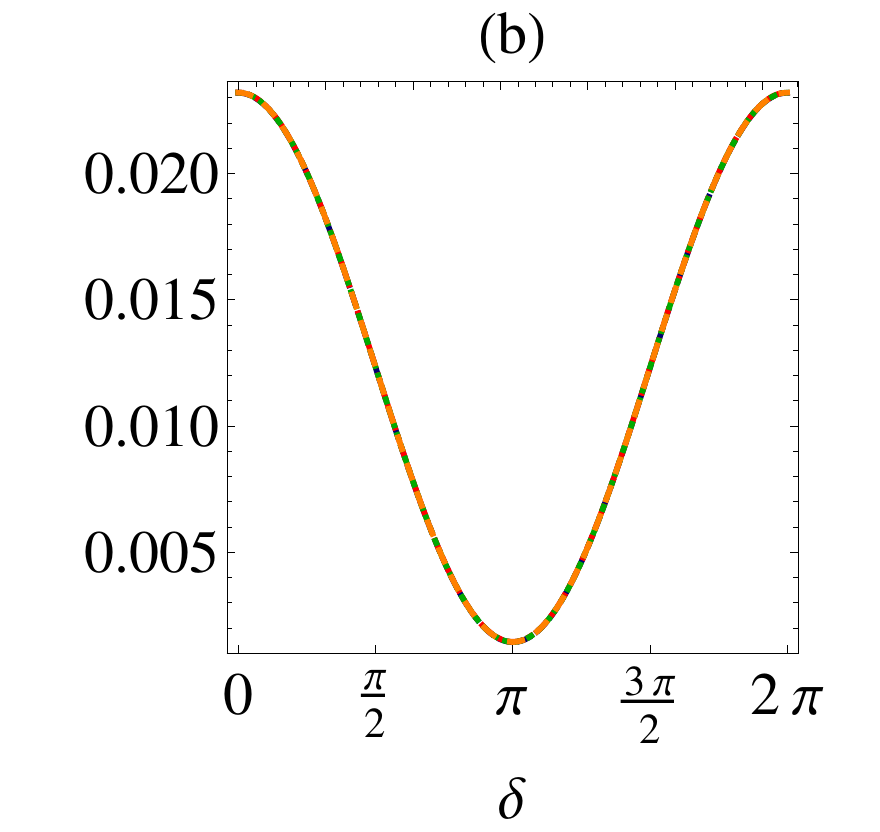}
  \caption{(Color online):  $\Delta\gamma$ vs the squeezing parameter $\delta$ for the different CS: even  (blue dashed line), odd  (red dotdashed line),  and Yuker-Stoler $\pm \pi/2$ (green dotted and orange medium dashed lines)  respectively. In Fig (a), $|\alpha|=0.5$; in (b) $ |\alpha| = 2$ and $r=1$.  }
\label{onerr}
 \end{minipage}
\end{figure}

We see in Fig. \ref{oner}(a) that we are able to distinguish between the three states when the amplitude of the CS is small (i.e $|\alpha|\leq 2$). However as shown in \ref{oner}(b), the three states become indistinguishable for $|\alpha| > 2$. We also show in Fig. \ref{onerr} the relationship $\Delta \gamma$ vs $\delta$ where $\delta$ is the phase of the squeezing parameter.Again we see that the three CS are indistinguishable for values of $|\alpha| \geq 2$ as shown in Fig. \ref{onerr}(b). To conclude, we note that distinguishing these cat states is best achieved with small intensities of light  and for small squeezing parameters.

\subsection{Relationship between $\Delta \gamma$ vs $\theta$ for the different cat states.}

Superposition of coherent states of light exhibit oscillations in their photon number distribution. By resolving these oscillations one is able to distinguish between the even and odd CS respectively.

\begin{figure}[h!]
 \begin{minipage}{.5\textwidth}  \includegraphics[width=.48\textwidth]{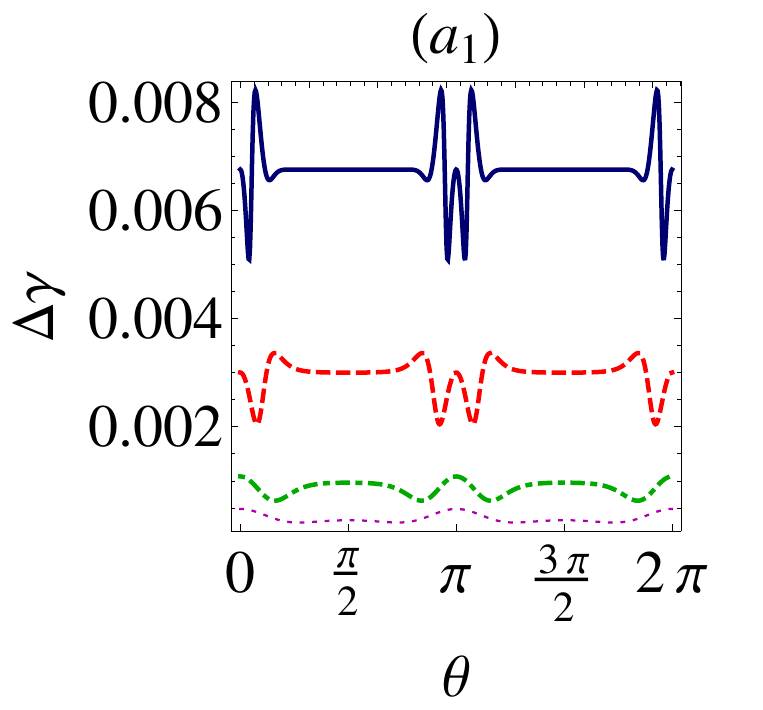} \includegraphics[width=.45\textwidth]{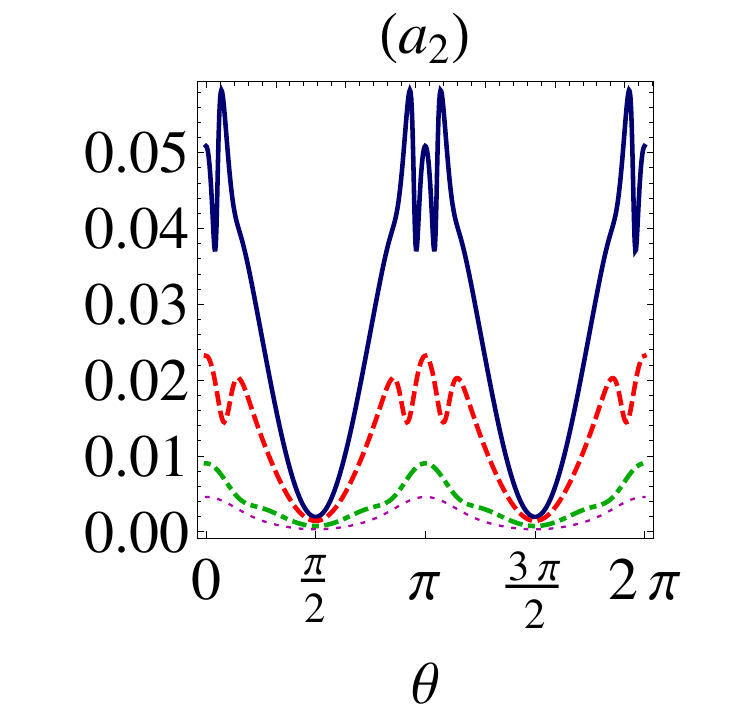}   
\label{theta}
 \end{minipage}
 \begin{minipage}{.5\textwidth}  \includegraphics[width=.48\textwidth]{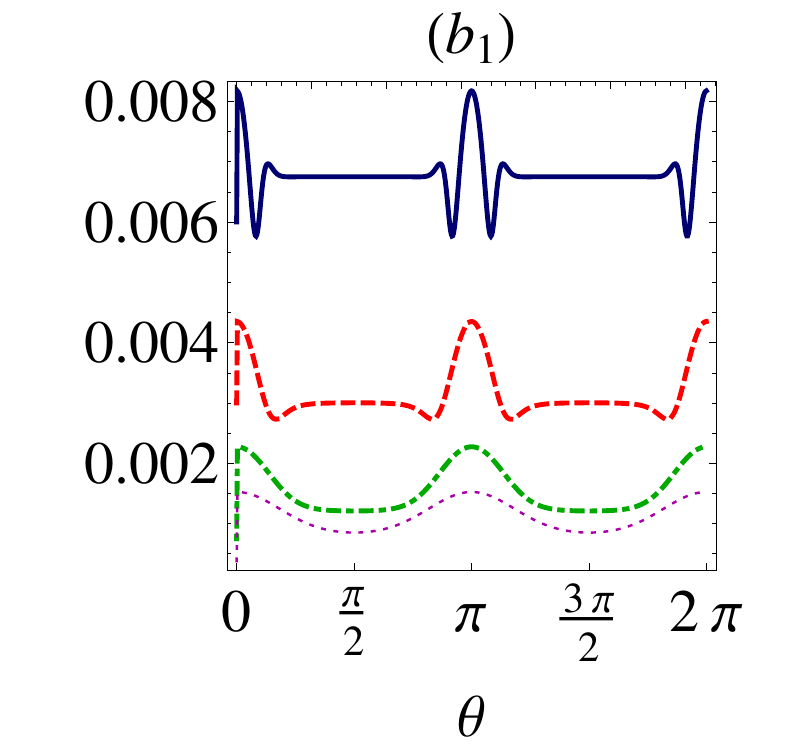} \includegraphics[width=.45\textwidth]{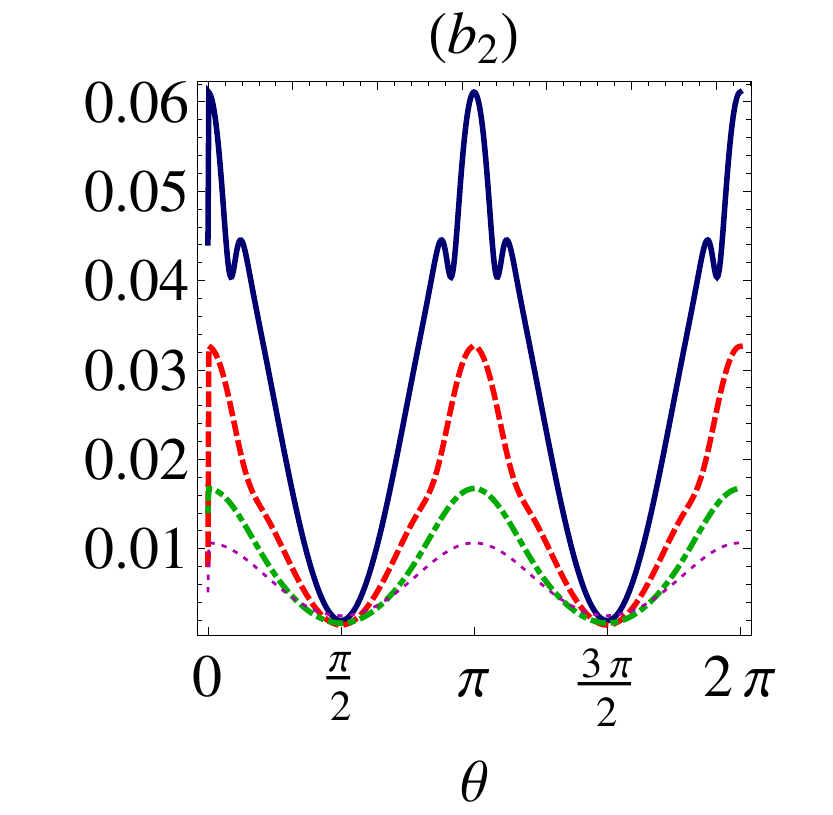}   
 \end{minipage}
 \begin{minipage}{.5\textwidth}  \includegraphics[width=.48\textwidth]{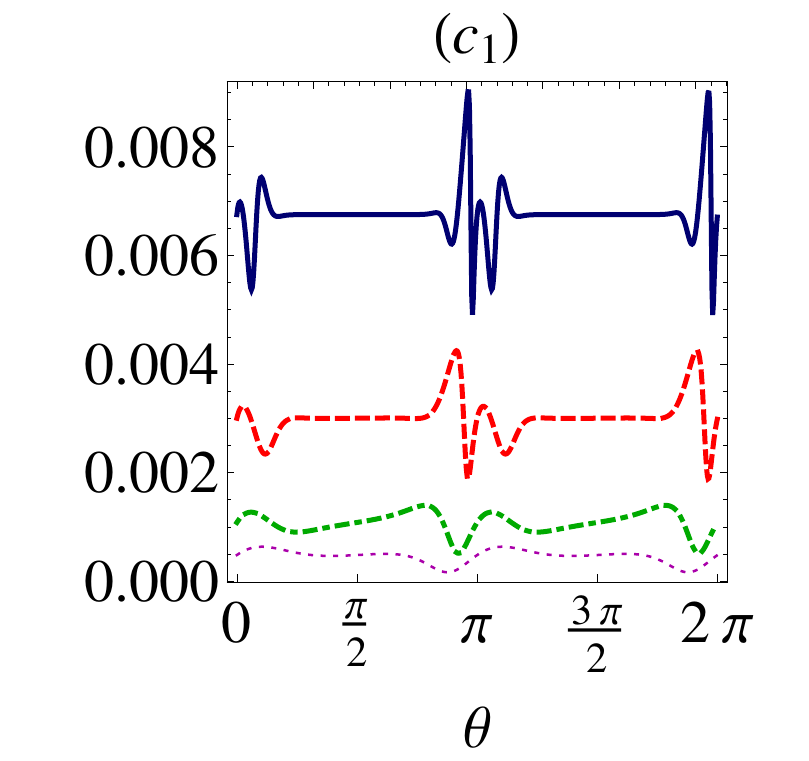} \includegraphics[width=.45\textwidth]{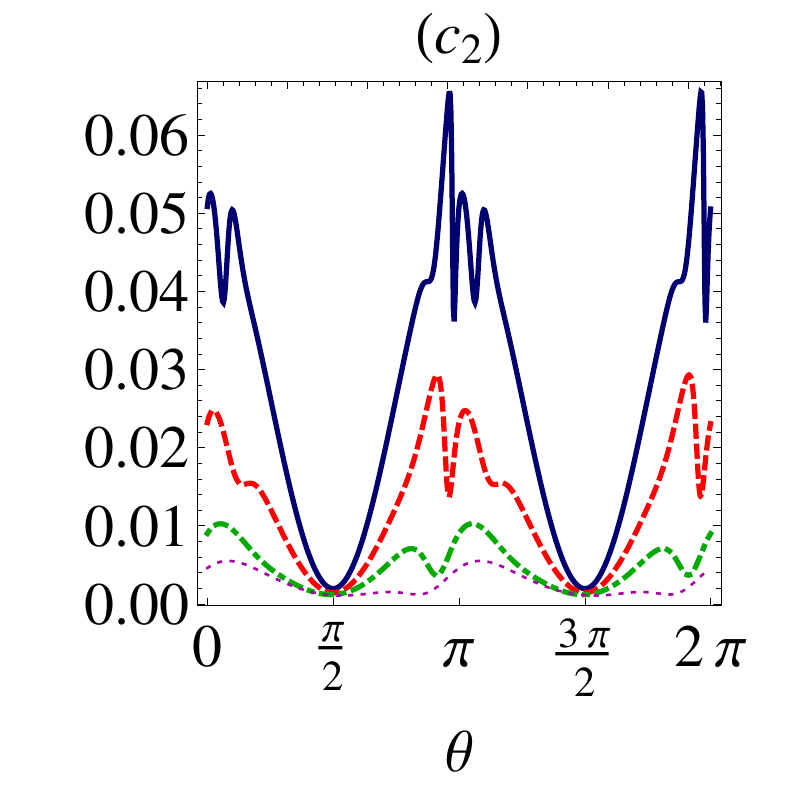}   \caption{(Color online):  Relative phase difference $\Delta\gamma$ vs the parameter $|\theta|$  for the different cat states: ($a_{i}$) even, ($b_{i}$) odd, ($c_{i}$) Yoker-Stoler $\pm$ CS   respectively ($i=1,2$) with $|\alpha|=3$ (solid blue line), $|\alpha|=2$ (red dashed line), $|\alpha|=1.2$ (green dotdashed line), $|\alpha|=0.8$ (magenta dotted line). The "1" subscript shows relationship fr zero squeezing and "2" subscript shows squeezing for $r=1$.}
\label{theta2}
 \end{minipage}
\end{figure}

We see in \eqref{result} that the interferometric phase difference $\Delta \gamma$ is a sensitive function of $\theta$--the phase of the component coherent states $\ket{\alpha}, \ket{-\alpha}$. Fig.~\ref{theta2} shows the relationship between $\Delta \gamma$ and $\theta$. We see distinct oscillatory behaviour for each of the three cases, but is damped by the term $\exp(-2| \alpha|^{2} \sin^{2}(\theta))$ resulting from the non-orthogonality of the coherent states. Note that in the case of zero squeezing, the oscillations in $\Delta \gamma$ only appear for values of $\theta$ for which $\langle \beta| \alpha \rangle \neq 0$; in other words when  the two coherent states are not distinguishable \cite{buzek,schleich}. On the other hand when squeezing is introduced, the oscillatory behaviour becomes more defined even though it still damped due to the non-orthogonality of the coherent states.

\subsection{Investigating the general Cat state}

\begin{figure}[h!]
 \begin{minipage}{.5\textwidth}  \includegraphics[width=.5\textwidth]{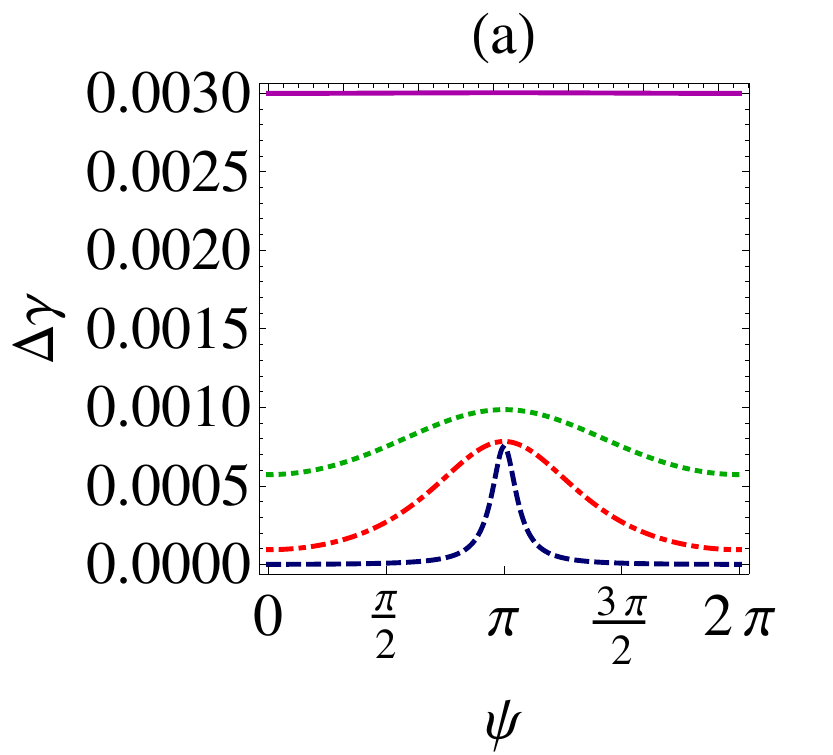} \includegraphics[width=.48\textwidth]{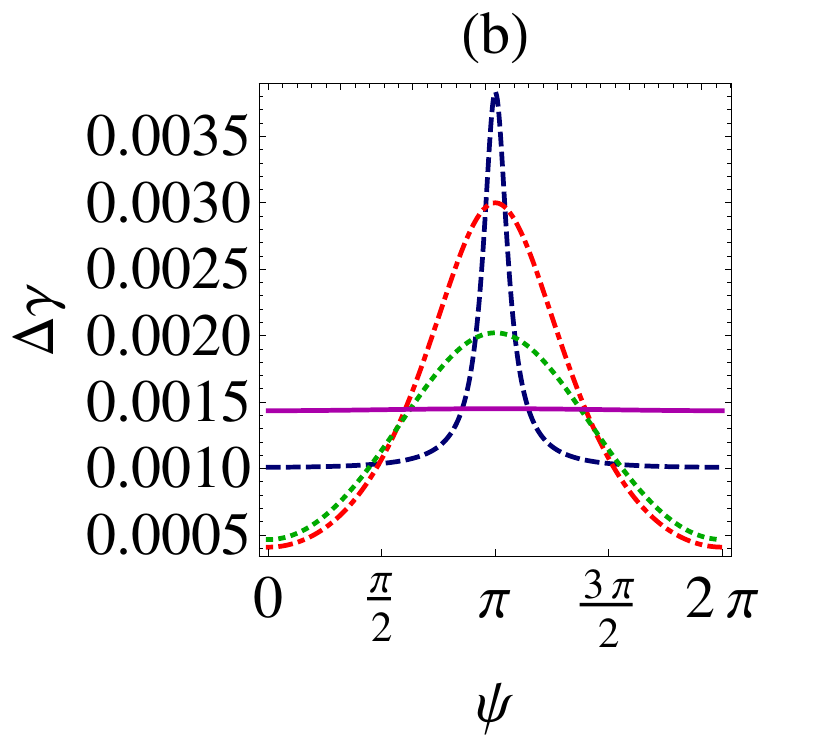}   
 \caption{(Color online):  Relative phase difference $\Delta\gamma$ vs the parameter $\zeta$  for the different values of $|\alpha|$: $|\alpha|=0.1 $ (Blue dashed line), $|\alpha|=0.6$ (red dotdashed line), $|\alpha|=1$ (green dotted lline), $|\alpha|=2$ (magenta solid line) and $|\zeta|=0$ for (a); $|\zeta|=1$ for (b). We see the curves approach a delta function for small values of $|\alpha|$}
\label{theta4}
 \end{minipage}
\end{figure}

We now turn from studying  the properties of the quantum state \eqref{rho} for $\psi=0,\pi, \pi/2$ to  investigating  the more general cat state \eqref{statesim}, for a given $\psi$. We represent the relationship $\Delta \gamma$ vs $\psi$ in Fig.~\ref{theta4}. For different values of $|\alpha|$ , we have Gaussian shapes that are centered at $\zeta = \pi$ and stretched by increasing $|\alpha|$. The Gaussian height increases for non-zero squeezing Fig.~\ref{theta4}(b).

\section{Discussion and Conclusion}
In this work we have shown that the mode-invisibility technique provides (at least in principle) a good measurement scheme for  observing the quantum nature of a superposition of coherent states.  We demonstrated this explicitly for three different Schroedinger cat states \eqref{fours}. For small values of the magnitude $\alpha$ of the coherent state parameter,  we find it straightforward to distinguish these states. The distinguishability of the three CS is enhanced by squeezing. 

To summarize, in contrast to the several ways in which the nonclassical properties of coherent states have been investigated, our method provides a measure for studying the behaviour of the superposed cat state, distinguishing between the CS most importantly in a non-destructive way.  Of course the natural question  is how to realize this mode-invisibility technique in the laboratory and use it to study the decoherence properties of these cat states. We leave this project for future study.

\section{Acknowledgments}

This work was supported in part by the Natural Sciences and Engineering Research Council of Canada.  Marvellous Onuma-Kalu thanks Kae Nemeto and her group at the National Institute of Information, Tokyo, Japan for valuable comments.

\bibliography{references}
 
\end{document}